\documentclass[12pt,a4paper]{article}
\usepackage{latexsym}
\def\be{\begin{equation}}
\def\ee{\end{equation}}
\def\bea{\begin{eqnarray}}
\def\eea{\end{eqnarray}}
\def\pd{\partial}

\def\al{\alpha}
\def\lm{\lambda}
\def\sig{\sigma}
\def\om{\omega}
\def\eps{\epsilon}

\def\vmu{\vec\mu}

\def\vnu{\vec\nu}
\def\vnup{\vec\nu'}
\def\vrho{\vec\rho}

\def\kp{\kappa}
\def\vkap{\vec\kappa}
\def\vkp{\vec\kappa'}
\def\vtau{\vec\tau}

\def\veta{\vec\eta}
\begin{document}
\begin{flushright}
DCPT/01/21
\end{flushright}
\vspace{0.2cm}
\begin{center}
\begin{Large}
{\bf Multi-Field Generalisations of the Klein-Gordon
Theory associated with $p$-Branes}
\end{Large}
\\
\vspace{0.5cm}
{\bf David B. Fairlie
$\footnote{e-mail: David.Fairlie@durham.ac.uk}$ 
and Tatsuya Ueno
$\footnote{e-mail: Tatsuya.Ueno@durham.ac.uk, \ 
ueno@funpth.phys.sci.osaka-u.ac.jp}$}\\
\vspace{0.3cm}
Department of Mathematical Sciences,\\
         Science Laboratories,\\
         University of Durham,\\
         Durham, DH1 3LE, England \\
\end{center}

\vspace{0.2cm}
\begin{abstract} 
The purpose of this article is to initiate a study of a class of 
Lorentz invariant, yet tractable, Lagrangian Field Theories which 
may be viewed as an extension of the Klein-Gordon Lagrangian to 
many scalar fields in a novel manner. 
These Lagrangians are quadratic in the Jacobians of the participating 
fields with respect to the base space co-ordinates. 
In the case of two fields, real valued solutions of the equations of 
motion are found and a phenomenon reminiscent of instanton behaviour 
is uncovered; an ansatz for a subsidiary equation which implies a 
solution of the full equations yields real solutions in 
three-dimensional Euclidean space.
Each of these is associated with a spherical harmonic function.
\end{abstract}

%%%%%%%%%%%%%%%%%%%%%%%
\section{Introduction}%
%%%%%%%%%%%%%%%%%%%%%%%
Recently a class of field theories which arise as a continuation 
of the Dirac-Born-Infeld theories to the case where the dimension 
of the base space is larger than that of the target space was 
investigated \cite{bf1,dbf1,tu}.
We named these theories Companion theories, as they appeared 
as associated with string and D-brane theories, and the 
corresponding Lagrangians as Companion Lagrangians,
which were of square-root form,
%-----------------------------------------------------------------
\be
{\cal L} = \sqrt{\det\left|\frac{\pd \phi^i}{\pd x^\mu}
\frac{\pd \phi^j}{\pd x^\mu}\right|} \ ,
\label{cl}
\ee
%-----------------------------------------------------------------
where the set of functions $\phi^i$ $(i=1,2,\cdots,p+1)$ each 
dependent upon co-ordinates $x^\mu$ ($\mu=1,2,\cdots,d > p+1$) 
is a mapping from the base space to the target space.
A remarkable property of the theory is that the equations of 
motion derived from ${\cal L}$ are invariant under arbitrary
field redefinitions $\phi^i \rightarrow \phi'^i(\phi^j)$.
In consequence, a large class of solutions may be found, and these 
equations are completely integrable when $d\,=\,p+2$. 
However the solutions are for the most part given in implicit form. 
\par

In this article, we shall instead study the Lagrangian density which is 
the square of (\ref{cl}), for which it will turn out that it inherits 
much of the integrability of the previous case, but permits solutions 
of the equations of motion in explicit form. 
Indeed, while this investigation was motivated by the notion of the 
Companion Lagrangian, it may be regarded as a study in its own right 
of a natural generalisation of the Klein-Gordon Lagrangian to many 
fields obtained by replacing $\pd \phi / \pd x_\mu$ by the Jacobian 
$J_{\mu_1\mu_2 \cdots \mu_{p+1}}$ of $p+1$ fields with respect 
to a set of the co-ordinates, and summing over all combinations of
$J^{\,2}_{\mu_1\mu_2 \cdots \mu_{p+1}}$.
\par

After a short description of the equations of motion for the general 
case of $p+1$ fields in $d$ dimensions, we analyse the specific
example of two fields.
We also give attention to a first order differential equation 
as an ansatz to solve the full equations of motion with $p+1$ 
fields in $d=p+2$ dimensional Euclidean space.
In the simplest, non-trivial example among such theories, 
corresponding to two fields in three dimensions, we explicitly 
demonstrate a large class of solutions, including a 
solution of the ansatz equation associated with each spherical
harmonic function.
\par

In this article, we shall restrict ourselves in the main to the 
massless case.
The example studied should not be confused despite analogous 
duality features with the study of antisymmetric tensor fields, 
as in Freedman and Townsend \cite{ft}.
Here the antisymmetric field has an additional structure, as it is 
constructed from two scalar fields. 

%%%%%%%%%%%%%%%%%%%%%%%%%%%%%%
\section{Companion Equations}%
%%%%%%%%%%%%%%%%%%%%%%%%%%%%%%
%%%%%%%%%%%%%%%%%%%%%%
\subsection{Notation}%
%%%%%%%%%%%%%%%%%%%%%%
We work in $d$-dimensional flat space with the totally 
antisymmetric tensor $\eps_{\nu_1\nu_2\ldots\nu_d}$ 
with $\eps_{12\ldots d}=+1$.
Indices with an arrow above them denote a set of several 
indices. 
$\vmu,\vnu,\vrho$ each have $p+1$ components, e.g., 
$\vnu=\{\nu_1,\nu_2,\ldots,\nu_{p+1}\}$. 
$\vtau,\vkap,\veta$ each have $(d-p-1)$ components, e.g., 
$\vkap = \{\kp_1, \kp_2,\ldots,\kp_{d-p-1}\}$.
If the prime $'$ is used for the indices, 
$\vmu,\vnu$ or $\vtau,\vkap$, then their components start 
from the second entry of un-primed ones, e.g., 
$\vnup=\{\nu_2,\ldots,\nu_{p+1}\}$ and 
$\vkp=\{\kp_2,\ldots,\kp_{d-p-1}\}$.
\par

%%%%%%%%%%%%%%%%%%%%%%%
\subsection{Jacobians}%
%%%%%%%%%%%%%%%%%%%%%%%
We define Jacobians ${\tilde J}_{\vkap}$ and their duals $J_{\vnu}$
as 
%---------------------------------------------------------------------
\bea
{\tilde J}_{\vkap}= {\tilde J}_{\kp_1\kp_2\ldots\kp_{d-p-1}}
&&=\eps_{\kappa_1\kappa_2\ldots\kappa_{d-p-1}\nu_1\nu_2\ldots\nu_{p+1}}
\phi^1_{\nu_1}\phi^2_{\nu_2}\ldots\phi^{p+1}_{\nu_{p+1}} 
\nonumber \\
&&= {1 \over (p+1)!}\, \eps_{\vkap\vnu}\, \eps_{i_1\ldots i_{p+1}}
\phi^{i_1}_{\nu_1}\ldots\phi^{i_{p+1}}_{\nu_{p+1}}
\equiv {1 \over (p+1)!} \, \eps_{\vkap\vnu} \, J_{\vnu} \ ,
\label{j}
\eea
%---------------------------------------------------------------------
where $\phi^i_\nu = \pd \phi^i/\pd x^\nu$.
The derivatives of the Jacobians are
%---------------------------------------------------------------------
\be
{\pd {\tilde J}_{\vkap} \over \pd \phi^i_{\mu}} 
= {1 \over p!} \, \eps_{\vkap\mu\vnup} \, \eps_{ii_2\ldots i_{p+1}}
\phi^{i_2}_{\nu_2}\ldots\phi^{i_{p+1}}_{\nu_{p+1}}
\equiv {1 \over p!} \, \eps_{\vkap\mu\vnup}\, J_{i,\vnup} 
\ .  \label{de1j}
\ee
%---------------------------------------------------------------------
Using the Jacobians, the Companion Lagrangian ${\cal L}_1$ without 
square-root is written as
%---------------------------------------------------------------------
\be
{\cal L}_1 \,=\, \det\left|\frac{\pd \phi^i}{\pd x^\mu}
\frac{\pd \phi^j}{\pd x^\mu}\right|
\,=\, {1 \over (d-p-1)!} \, {\tilde J}_{\vkap} 
{\tilde J}_{\vkap}
\,=\,{1 \over (p+1)!} \, J_{\vmu} J_{\vmu} \ .
\label{cl2}
\ee
%---------------------------------------------------------------------
The equation of motion for ${\cal L}_1$ is then given by
%---------------------------------------------------------------------
\be
\pd_\mu \left({\pd {\cal L}_1 \over \pd \phi^i_\mu}\right) 
= {2 \over (d-p-1)!} \, {\pd {\tilde J}_{\vkap} \over \pd \phi^i_\mu}
{\pd {\tilde J}_{\vkap} \over \pd \phi^j_\nu} \, \phi^j_{\mu\nu}
= {2 \over p!} \, J_{i, \vnup} \, \pd_\mu J_{\mu \vnup} = 0 \ .
\label{gem}
\ee
%---------------------------------------------------------------------

%%%%%%%%%%%%%%%%%%%%%%%%%%%%%%%%%%%%%%%%%%%%%%%%%%%%%
\section{Solutions of Quadratic Jacobian Lagrangian}%
%%%%%%%%%%%%%%%%%%%%%%%%%%%%%%%%%%%%%%%%%%%%%%%%%%%%%
\label{sqjl}
Let us consider the specific case of two fields, in which the 
Jacobians are given, with $\phi^1=u$ and $\phi^2=v$,
%---------------------------------------------------------------------
\be
J_{\mu\nu}\,=\,\frac{\pd u}{\pd x^\mu}\frac{\pd v}{\pd x^\nu}\,
-\,\frac{\pd u}{\pd x^\nu} \frac{\pd v}{\pd x^\mu} \ .
\label{j2}
\ee
%---------------------------------------------------------------------
It is noteworthy that $J_{\mu\nu}$ is like a field strength 
$F_{\mu\nu}$ for the gauge potential 
%---------------------------------------------------------------------
\be
A_\mu\,=\, u\frac{\pd v}{\pd x^\mu}+\frac{\pd w}{\pd x^\mu} \ ,
\ee
%---------------------------------------------------------------------
where the field $w$ represents an ambiguity by gauge 
transformation.
This is the Clebsch representation for a vector field in the case of 
three dimensions \cite{dbf1, cos}.
Using subscripts to denote derivatives, e.g.
$u_\mu = \pd u/\pd x^\mu$,
the equations of motion (\ref{gem}) in this case are simply 
%---------------------------------------------------------------------
\be 
u_\nu \frac{\pd}{\pd  x^\mu}J_{\mu\nu}\,=\,0 \ , 
\qquad  v_\nu \frac{\pd}{\pd  x^\mu}J_{\mu\nu}\,=\, 0 \ .
\label{2em}
\ee
%---------------------------------------------------------------------
Note that these equations mean that the vector 
$\pd J_{\mu\nu}/\pd x^\mu$ is orthogonal to both the vectors $u_\nu$ 
and $v_\nu$, and this implies the following equation in three dimensions,
%---------------------------------------------------------------------
\be
\frac{\pd}{\pd  x^\mu}(u_\mu v_\nu-u_\nu v_\mu) 
\,=\,\lm(u,v) \, \eps_{\nu\rho\sigma}u_\rho v_\sigma \ ,
\label{lambda}
\ee
%---------------------------------------------------------------------
where $\lm$ is a function of $u,\ v$.
\par

There is a large class of solutions which may be simply  
categorised as follows,
%---------------------------------------------------------------------
\be
u\,=\,f(a_\rho x^\rho,\ b_\rho x^\rho) \ , \qquad 
v\,=\,g(a_\rho x^\rho,\ b_\rho x^\rho) \ ,
\label{sol}
\ee
%---------------------------------------------------------------------
where $f,\ g$ are arbitrary functions of two variables and the 
constant vectors $a_\rho,\ b_\rho$ satisfy
%---------------------------------------------------------------------
\be 
(a_\rho a_\rho)(b_\sigma b_\sigma)\,-\,(a_\rho b_\rho)^2 \,=\, 0
\ .
\label{cond}
\ee
%---------------------------------------------------------------------
Notice that in Euclidean space such solutions, if they are to be 
nontrivial, are necessarily complex. 
The trivial solution $a_\rho\,=\,\lambda b_\rho$ implies that 
$u$ and $v$ are functionally related. 
This result is a direct generalisation of the Klein-Gordon situation, 
that $\phi(\vec a\cdot\vec x)$ is a solution of the massless equation 
iff $\vec a$ is null. 
A specific example of this form of solution in three dimensions is 
given by
%---------------------------------------------------------------------
\be
u\,=\,f(x+i\sin\theta y,z+i\cos\theta y) \ , \qquad
v\,=\,g(x+i\sin\theta y,z+i\cos\theta y) \ .
\label{soln}
\ee
%---------------------------------------------------------------------
In Minkowski space, of course, real solutions may be obtained by 
setting $t=iy$. 
In fact a yet more general construction can be given; in three 
dimensions it is
%---------------------------------------------------------------------
\be
u\,=\,F(a_\rho x^\rho,\ b_\rho x^\rho,\ c_\rho x^\rho) \ , \qquad
v\,=\,G(a_\rho x^\rho,\ b_\rho x^\rho,\ c_\rho x^\rho) \ ,
\label{sol2}
\ee
%---------------------------------------------------------------------
where $F,\ G$ are arbitrary functions of three variables and the 
constant vectors $a_\rho,\ b_\rho,\ c_\rho$ satisfy
%---------------------------------------------------------------------
\be 
(a_\rho a_\rho)(b_\sigma b_\sigma)\,-\,(a_\rho b_\rho)^2 \,=\, 0
\ , \qquad 
c_\rho\,=\, \eps_{\rho\tau\sigma}a_\tau  b_\sigma \ ,
\label{cond1}
\ee
%---------------------------------------------------------------------
while in higher dimensions the definition of $c_\rho$ is replaced 
by
%---------------------------------------------------------------------
\be
c_\rho\,=\,\eps_{\rho\lambda_1\lambda_2\dots\tau\sigma}\, 
k^1_{\lambda_1} k^2_{\lambda_2}\dots a_\tau \, b_\sigma \ ,
\label{cond2}
\ee
%---------------------------------------------------------------------
and $\vec k^1\ \vec k^2\dots$ are constant vectors orthogonal
to $\vec a$ and $\vec b$.
There is an alternative strategy in seeking solutions, which we 
elucidate below.

%%%%%%%%%%%%%%%%%%%%%%%%%%%%%%%%%%%%%%%%%
\section{Ansatz for Companion Equations}%
%%%%%%%%%%%%%%%%%%%%%%%%%%%%%%%%%%%%%%%%%
\label{afce}
In this section, we consider $p+1$ fields in $p+2$ dimensional
Euclidean space and obtain auxiliary equations whose solution 
implies that of the full equations of motion in the 
manner of the equations of self-duality. 

%%%%%%%%%%%%%%%%%%%%%%%%%%%%%%%%%%%%%%%%%%%%%%%%
\subsection{Ansatz equations for $(d=p+2,p+1)$}%
%%%%%%%%%%%%%%%%%%%%%%%%%%%%%%%%%%%%%%%%%%%%%%%%
In the $d=p+2$ dimensional case, the Jacobian defined in 
(\ref{j}) has one index $\kp$; ${\tilde J}_\kp = {1 \over (p+1)!}
\eps_{\kp \nu_1 \ldots \nu_{p+1}} J_{\nu_1 \ldots \nu_{p+1}}$.
We make the assumption  that the Jacobian is expressed
as the derivative of a function $G(x)$,
%---------------------------------------------------------------------
\be
{\tilde J}_\kp \,=\, {1 \over (p+1)!} \, \eps_{\kp \nu_1 \ldots 
\nu_{p+1}} J_{\nu_1 \ldots \nu_{p+1}} \,=\,
{1 \over (p+1)!}\, \eps_{\kp \vnu} \, J_{\vnu}
\,=\, \pd_\kp G(x) \ ,
\label{an-1}
\ee
%---------------------------------------------------------------------
or equivalently, in terms of the dual $J_{\vnu}$,
%---------------------------------------------------------------------
\be
J_{\vnu} = \eps_{\kp \vnu} \, \pd_\kp G(x) \ .
\label{an-2}
\ee
%---------------------------------------------------------------------
{}From (\ref{an-2}), we can easily see solutions of the ansatz 
to satisfy the Companion equation of motion (\ref{gem}),
since $\pd_{\nu_1}J_{\nu_1 \vnup}=\eps_{\kp \nu_1\vnup}
\pd_{\nu_1} \pd_\kp G(x) = 0$.
Differentiating the LHS of (\ref{an-1}) and using the Jacobi 
identity for $J_{\vnu}$, $\pd_{[\kp} J_{\vnu]} = 0$, 
we obtain the harmonic equation for $G(x)$,
%---------------------------------------------------------------------
\be
\Box G(x) = \pd_\mu \pd_\mu G(x) = 0 \ .
\label{he}
\ee
%---------------------------------------------------------------------
The relation (\ref{an-2}) between $J_{\vnu}$ and $\pd_\kp G(x)$ 
may be considered analogous to the implications of self-duality in 
monopole theory. 
The Jacobi identity for the former gives the harmonic equation
for the latter, while the integrability condition for the latter 
leads to a divergence-free equation for the former which solves the 
Companion equations (\ref{gem}). 
\par

%%%%%%%%%%%%%%%%%%%%%%%%%%%%%%%
\subsection{$(d=3, p=1)$ case}%
%%%%%%%%%%%%%%%%%%%%%%%%%%%%%%%
Let us apply the ansatz equation (\ref{an-1}) for the case
with two fields in three dimensions,
%------------------------------------------------------------------------
\be
{\tilde J}_\mu \,=\, {1 \over 2} \, \eps_{\mu \nu \sig}
J_{\nu \sig} \,=\, \eps_{\mu \nu \sig} \, u_\nu \, v_\sig 
\,=\, \pd_\mu G(x) \ .
\label{an-132}
\ee
%------------------------------------------------------------------------
In \cite{cosmas}, this equation has appeared as a dual version 
of the Poisson Bracket version of the three-dimensional Nahm 
equations, for which Ward has given an implicit solution 
\cite{ward}. 
{}From (\ref{an-132}), we see that the vector $\pd_\mu G$ is 
orthogonal to both vectors $u_\nu$ and $v_\sig$.
As is well known, with the use of polar co-ordinates
%------------------------------------------------------------------------
\be
x = r \sin{\theta}\cos{\phi} \ , \quad
y = r \sin{\theta}\sin{\phi} \ , \quad
z = r \cos{\theta} \ ,
\label{polar}
\ee
%------------------------------------------------------------------------
the 3D harmonic equation for $G$ is solved in terms of spherical 
harmonic functions,
%------------------------------------------------------------------------
\bea
&&G = \sum_{l=0}^{\infty} \sum_{m=-l}^l \, a_l^{\, m} \, 
G_l^{\, m} \ , \qquad G_l^{\, m} = {1 \over r^{l+1}} \, 
Y_l^m (\theta,\phi) \ , \nonumber \\
&&Y_l^m (\theta,\phi) = P_l^{|m|}(\cos{\theta}) \, e^{i m \phi} \ ,
\label{Y}
\eea
%------------------------------------------------------------------------
where $a_l^{\, m}$ are constants and $P_l^{|m|}(\cos{\theta})$ are 
associated Legendre polynomials.
\par

Let us first apply each $G_l^{\, m}$ to the ansatz equation 
and find corresponding solutions of $u$ and $v$.
The homogeneity of the RHS of (\ref{an-132}) is of degree 
$-(l+2)$, as $G_l^{\, m}$ is of degree $-(l+1)$, 
Thus, at a first attempt to solve the ansatz, it is natural to 
impose a homogeneity condition upon both functions 
$u$ and $v$,
%------------------------------------------------------------------------
\be
u(tx,ty,tz) = t^{-k} \, u(x,y,z) \ , \qquad 
v(tx,ty,tz) = t^{-n} \, v(x,y,z) \ ,
\label{hom1}
\ee
%------------------------------------------------------------------------
or equivalently,
%------------------------------------------------------------------------
\be
x^\mu \pd_\mu u(x,y,z) = -k \, u(x,y,z) \ , \qquad 
x^\mu \pd_\mu v(x,y,z) = -n \, v(x,y,z) \  .
\label{hom2}
\ee
%------------------------------------------------------------------------
When $u$ and $v$ are expressed in terms of $\{r,\theta,\phi\}$,
their $r$ dependences are determined by the condition (\ref{hom1})
as $u = r^{-k} U(\theta,\phi)$ and $v = r^{-n} V(\theta,\phi)$.
The degrees of $u$, $v$ and $G$ are obviously related by
%------------------------------------------------------------------------
\be
k + n = l \ .
\label{degree}
\ee
%------------------------------------------------------------------------
We suppose the numbers $k$ and $n$ to be integers and also
$k \geq n \geq 0$ to make sure  that $u$ and
$v$ are well behaved as $r \rightarrow \infty$. 
\par 

In the $l=0$ case, $G = G_0^{\, 0} = 1/r$ and functions $u$ and 
$v$ are homogeneous of degree zero.
Then a solution of the ansatz (\ref{an-132}) is easily obtained as
%------------------------------------------------------------------------
\be
u = \cos{\theta} = {z \over r} \ , \qquad
v = \phi = \arctan{y \over x} \ .
\label{sps}
\ee
%------------------------------------------------------------------------
\par

For the $l>0$ case, multiplying (\ref{an-132}) by 
$\eps_{\mu\lm\rho}x_\lm$, with the use of the homogeneity 
condition of $u$ and $v$, we have
%------------------------------------------------------------------------
\be
- k \, u \, v_\rho + n \, v \, u_\rho = \eps_{\mu\lm\rho}x_\lm G_\mu \ .
\label{an-an}
\ee
%------------------------------------------------------------------------
Conversely, acting with the operator $\eps_{\tau\sig\rho}\pd_\sig$
upon (\ref{an-an}), we have $l (\eps_{\tau\sig\rho}u_\sig v_\rho 
- G_\tau) = x_\tau \triangle G$, showing the equivalence between
(\ref{an-132}) and (\ref{an-an}), provided $G(x)$ is a solution of 
the harmonic equation.\\
Let us consider the canonical transformation,
%------------------------------------------------------------------------
\be
u' \,=\, - {1 \over l} u \, v \ , \qquad 
v' \,=\, \ln{v^k \over u^n} \ , \qquad \{u',v'\}_{(u,v)} = 1 \ ,
\label{ct}
\ee
%------------------------------------------------------------------------
under which the ansatz equation (\ref{an-132}) is invariant.
Then (\ref{an-an}) is transformed into the form of the equation 
(\ref{an-an}) with $(k,n) = (l,0)$.
Therefore, in the following, we concentrate on the $(l,0)$-type 
solutions.
\par

For the $(l,0)$ case, the equation (\ref{an-an}) becomes
%------------------------------------------------------------------------
\be
v_\mu = \al(x) \, \eps_{\mu\lm\rho}x_\lm G_\rho = l_\mu G \ , 
\qquad 
\al(x) = {1 \over l \, u} \ ,
\label{V}
\ee
%------------------------------------------------------------------------
where $l_\mu$ are angular momentum operators, which are
rewritten in terms of $\{r,\theta,\phi \}$,
%------------------------------------------------------------------------
\be
l_1 = - \sin{\phi} {\pd  \over \pd \theta} 
- \cot{\theta} \cos{\phi} {\pd \over \pd \phi} \ , \quad
l_2 = \cos{\phi} {\pd \over \pd \theta} 
- \cot{\theta} \sin{\phi} {\pd \over \pd \phi} \ , \quad
l_3 = {\pd \over \pd \phi} \ .
\label{am}
\ee
%------------------------------------------------------------------------
The first equation in (\ref{V}) can be recast as
%------------------------------------------------------------------------
\be
v_\theta = - {\al(x) \, r \over \sin{\theta}}\, G_\phi \ , \qquad
v_\phi = \al(x) \, r \sin{\theta}\, G_\theta \ .
\label{V2}
\ee
%------------------------------------------------------------------------
The integrability condition 
$\pd_\phi v_\theta = \pd_\theta v_\phi$ gives a first order
equation for $\al(x)$,
%------------------------------------------------------------------------
\be
{G_\phi \over \sin^2{\theta}} \, \pd_\phi \ln{\al} 
+ G_\theta \, \pd_\theta \ln{\al} = - {1 \over \sin^2{\theta}}
(\sin{\theta} \, \pd_\theta (\, \sin{\theta} G_\theta) + G_{\phi 
\phi})
= - {\vec l}^{\, 2} \, G \ ,
\label{int}
\ee
%------------------------------------------------------------------------
where the operator ${\vec l}^{\, 2} = l_1^2 + l_2^2 + l_3^2$.
\par

For $G = G_l^{\, m}$ $(l=1,2,\cdots, m = 0,1,\cdots,l)$ in (\ref{Y}), 
the operators ${\vec l}^{\, 2}$ and $\pd_\phi = l_3$ give eigenvalues 
$-l(l+1)$ and $i m$, respectively.
Then (\ref{int}) becomes, with $q= \cos{\theta}$
%------------------------------------------------------------------------
\be
i m {P_l^m (q) \over \sin^2{\theta}} \, \pd_\phi \ln{\al} 
+ \sin^2{\theta} \, {d P_l^m (q)\over d q} \, \pd_q \ln{\al} 
= l(l+1) \, P_l^m (q) \ .
\label{int-2}
\ee
%------------------------------------------------------------------------
To find real solutions for $\al(x)$, we assume $\al(x) = \al(r,q)$,
which gives
%------------------------------------------------------------------------
\be
\pd_q \ln{\al(r,q)} = {l (l+1) P_l^m \over (1-q^2) (d P_l^m /d q)} 
\ .
\label{int-3}
\ee
%------------------------------------------------------------------------
The function $u$ is then given as
%------------------------------------------------------------------------
\be
u = {1 \over l \al} = { e^{-T(q)} \over l r^l} \ , \qquad
T(q) = l(l+1) \int dq \, {P_l^m (q) \over (1-q^2) d P_l^m (q) /dq}
\ .
\label{u}
\ee
%------------------------------------------------------------------------
The form of $v$ is easily evaluated by the second equation of 
(\ref{V2}), for $m \not= 0$,
%------------------------------------------------------------------------
\be
v = {i \over m} \, e^{T(q)} \, (1-q^2) {d P_l^m (q) \over d q} \, 
e^{i m \phi} \ .
\label{v}
\ee
%------------------------------------------------------------------------
We can easily check this solution to satisfy the first equation 
for $v_\theta$ in (\ref{V2}) by using the fact that the associated
Legendre polynomials satisfy
%------------------------------------------------------------------------
\be
{d \over d q}((1-q^2) {d \over d q}P_l^m (q) ) 
+ (l(l+1) - {m^2 \over 1-q^2}) P_l^m (q) = 0 \ .
\label{lp}
\ee
%------------------------------------------------------------------------
Having chosen the function $u$ to be real, we obtain
solutions for $G_l^{\, -m}$ $(m=1,\cdots,l)$
by taking the complex conjugation of (\ref{v}),
%------------------------------------------------------------------------
\be
u = {e^{-T(q)} \over l r^l} \ , \qquad
v = - {i \over m} \, e^{T(q)} \, (1-q^2) {d P_l^m (q) \over d q} \, 
e^{- i m \phi} \ .
\label{-m}
\ee
%------------------------------------------------------------------------
Taking a linear combination of solutions for $G_l^{\, m}$ and
$G_l^{\, -m}$, we have real solutions for the ansatz equation,
%------------------------------------------------------------------------
\bea
&& G = {P_l^m (q) \over r^{l+1}} (a \cos{m\phi} 
+ b \sin{m \phi}) \ , \quad \ 
(m=1,\cdots,l) \nonumber \\
&&u = {e^{-T(q)} \over l r^l} \, , \ \
v = {e^{T(q)} \over m} \, (1-q^2) {d P_l^m (q) \over d q}
(-a \sin{m\phi} + b \cos{m\phi}) \, ,
\label{real}
\eea
%------------------------------------------------------------------------
where $a$, $b$ are arbitrary constants.
\par

For $m=0$, with the use of the differential equation (\ref{lp}), 
(\ref{int-3}) becomes
%------------------------------------------------------------------------
\be
 \pd_q \ln{\al(r,q)} = - {d \over dq} 
 \ln{((1-q^2) {d P_l \over d q})} \ ,
\label{m=0-1}
\ee
%------------------------------------------------------------------------
which gives
%------------------------------------------------------------------------
\be
u = - {(1 - q^2) (d P_l /d q) \over l r^l} \ , \qquad
v = \phi \ .
\label{m=0-2}
\ee
%------------------------------------------------------------------------
Since the function $v$ does not depend upon the parameter $l$,
the sum of any number of $u$'s with different $l$'s also becomes
a solution of (\ref{V}) with the corresponding sum of 
$G_l^{\, 0} = P_l (q) /r^{l+1}$.
Indeed this solution satisfies the ansatz equation 
(\ref{an-132}), although it contains terms with different 
homogeneities.
This class of solutions of the ansatz equation is also obtained 
by taking $z$-derivatives of the $l=0$ solution (\ref{sps}),
%------------------------------------------------------------------------
\be
G = \sum_n a_n \, \pd_z^n {1 \over r} \ , \quad
u = \sum_n a_n \, \pd_z^n {z \over r} \ ,  \quad
v = \phi = \arctan{y \over x} \ ,
\label{sps-z}
\ee
%------------------------------------------------------------------------
where $a_n$ are constants.
These solutions also belong to the space of $m=0$ solutions. 
\par

To find an explicit form of $T(q)$ in (\ref{u}) for $m \not=0$ 
cases, let us introduce the following identity,
%------------------------------------------------------------------------
\be
(1-q^2) {d P_l^m \over d q} = (l+m) P_{l-1}^m - l q P_l^m
\ ,
\label{P-id1}
\ee
%------------------------------------------------------------------------
and a series expansion formula for $P_l^m$,
%------------------------------------------------------------------------
\bea
P_l^m (q) &=& (1-q^2)^{m \over 2} {d^m \over d q^m} P_l (q) 
\nonumber \\
&=& (1-q^2)^{m \over 2} \sum_{r=0}^{[(l-m)/2]}
{(-1)^r \over 2^r \, r!} {(2l-2r-1)!! \over (l-m-2r)!} \, q^{l-m-2 r}
\ . \label{P-id2}
\eea
%------------------------------------------------------------------------
Using these formulae, we can rewrite the integrand of 
$T(q)$ as a ratio of polynomials with respect to $q$,
%------------------------------------------------------------------------
\be
t(q) = {l(l+1) \, P_l^m \over (1-q^2) d P_l^m/dq} = 
{l(l+1) \, P_l^m (q) \over (l+m)P_{l-1}^m - l q P_l^m} \ .
\label{tq}
\ee
%------------------------------------------------------------------------
Note that the factor $(1-q^2)^{m \over 2}$ in the RHS of 
(\ref{P-id2}) does not appear in the above ratio.
The integral may be performed in principle by resolving the 
integrand into partial fractions. 
\par

\vspace{0.2cm}
In the following, we demonstrate explicit solutions of the 
ansatz equation (\ref{an-132}) for $m=l$ to $m=l-4$.
\par

\vspace{0.5cm} \noindent 
[1] $m=l$
%------------------------------------------------------------------------
\be
G =  {(\sin{\theta})^l \over r^{l+1}} \, e^{i l \phi} \ , \quad \ \ 
u = {(\cos{\theta})^{l+1} \over l \, r^l} \ , \quad
v = - i (\tan{\theta})^l  \, e^{i l \phi} \ .
\label{m=l}
\ee
%------------------------------------------------------------------------
\par

\vspace{0.5cm} \noindent
[2] $m=l-1$
%------------------------------------------------------------------------
\bea
G &=& {(\sin{\theta})^{l-1} \cos{\theta} \over r^{l+1}} \, e^{i(l-1)\phi} 
\ , \nonumber \\
u &=& {(l \cos^2{\theta} - 1)^{l+1 \over 2} \over l \, r^l} \ , \quad
v = - {i \over (l-1)} {(\sin{\theta})^{l-1} \over 
(l \cos^2{\theta} - 1)^{l-1 \over 2}}\, e^{i(l-1) \phi} \ .
\label{m=l-1}
\eea
%------------------------------------------------------------------------
\par

\vspace{0.5cm} \noindent
[3] $m=l-2$
%------------------------------------------------------------------------
\bea
G &=& {(\sin{\theta})^{l-2}((2l-1) \cos^2{\theta} -1) \over r^{l+1}}
\, e^{i(l-2)\phi} \ , \nonumber \\
u &=& {1 \over l \, r^l} (\cos{\theta})^{l(l+1) \over 5l-4}
((2l-1)l \cos^2{\theta} - (5l-4))^{2{(l+1)(l-1) \over 5l-4}} \ , \\
v &=& - {i \over (l-2)} \, {(\sin{\theta})^{l-2} 
(\cos{\theta})^{- {(l-2)^2 \over 5l-4}} \over 
((2l-1)l \cos^2{\theta} - (5l-4))^{(2l-1)(l-2) \over 5l-4}} \,
e^{i(l-2)\phi} \ . \nonumber 
\label{m=l-2}
\eea
%------------------------------------------------------------------------
\par

\vspace{0.5cm} \noindent
[4] $m=l-3$
%------------------------------------------------------------------------
\bea
G &=& {(\sin{\theta})^{l-3}\cos{\theta}((2l-1) \cos^2{\theta} - 3) 
\over r^{l+1}}\, e^{i(l-3)\phi} \ , \nonumber \\
u &=& {1 \over l \, r^l} 
{(\om +9(l-1) - 2l(2l-1)\cos^2{\theta})^
{(l+1) {3(l-3) + \om \over 4 \om}} \over 
(\om -9(l-1) + 2l(2l-1)\cos^2{\theta})^
{(l+1) {3(l-3) - \om \over 4 \om}}} \ , \\
v &=& i{e^{i(l-3)\phi} (\sin{\theta})^{l-3} \over 4 l(l-3)(2l-1)}
{(\om -9(l-1) + 2l(2l-1)\cos^2{\theta})^
{(l-3) {3(l+1) - \om \over 4 \om}}
\over (\om +9(l-1) - 2l(2l-1)\cos^2{\theta})^
{(l-3) {3(l+1) + \om \over 4 \om}}} \ , \nonumber 
\label{m=l-3}
\eea
%------------------------------------------------------------------------
where $\om = \sqrt{3(19 l^2 - 50 l + 27)}$\, .
\par

\vspace{0.5cm} \noindent
[5] $m=l-4$
%------------------------------------------------------------------------
\bea
G &=& {(\sin{\theta})^{l-4}((2l-1)(2l-3) \cos^4{\theta} 
- 6 (2l-3) \cos^2{\theta}+3) \over r^{l+1}} \, e^{i(l-4)\phi} \ ,
\nonumber \\
u &=& {1 \over l \, r^l} (\cos{\theta})^{l(l+1) \over 9l-16}
{(l(2l-1) \cos^2{\theta} - (7l-8) - 2 \xi)^{2(l+1)(l-2)(l-4 + \xi) 
\over (9l-16) \xi} \over (l(2l-1) \cos^2{\theta} -(7l-8) + 2 \xi)^
{2 (l+1)(l-2) (l-4 - \xi) \over (9l-16) \xi}} \ , \\
v &=& -i {(2l-3) \over l(l-4)(2l-1)} \, (\sin{\theta})^{l-4}
(\cos{\theta})^{-{(l-4)^2 \over 9l-16}} \times 
\nonumber \\
&& \times {(l(2l-1)\cos^2{\theta} - (7l-8) +2 \xi)^
{{2(l+1)(l-2)(l-4-\xi) \over (9l-16)\xi}+1} \over 
(l(2l-1)\cos^2{\theta} - (7l-8) - 2 \xi)^
{{2(l+1)(l-2)(l-4+\xi) \over (9l-16)\xi}-1}} \, e^{i(l-4)\phi} 
\nonumber \ ,
\label{m=l-4}
\eea
%------------------------------------------------------------------------
where $\xi = \sqrt{(l-2)(11 l^2 - 40 l +24)/(2l-3)}$\, .
\par

%%%%%%%%%%%%%%%%%%%%%%%%%%%%%%%%%
\section{Fourier mode solutions}%
%%%%%%%%%%%%%%%%%%%%%%%%%%%%%%%%%
The equations of motion (\ref{gem}) in the case of two fields can be
recast as, in general dimension,
%------------------------------------------------------------------------
\bea
\left(\frac{\pd v}{\pd x^\mu}\frac{\pd}{\pd x^\nu}
-\frac{\pd v}{\pd x^\nu}\frac{\pd}{\pd x^\mu}\right)^2u\,=\,0 
\ , \label{feq1}\\
\left(\frac{\pd u}{\pd x^\mu}\frac{\pd}{\pd x^\nu}
-\frac{\pd u}{\pd x^\nu}\frac{\pd}{\pd x^\mu}\right)^2v\,=\,0 
\ . \label{feq2}
\eea
%------------------------------------------------------------------------
Suppose now we look for a solution of the form
%------------------------------------------------------------------------
\be 
u\,=\, A\exp(k_\nu x^\nu) \ ,\ \ v\,=\, B\exp(p_\nu x^\nu)\,+\, 
C\exp(q_\nu x^\nu) \ , \label{ann}
\ee
%------------------------------------------------------------------------
where $A,\ B,\ C$ are constants, and $\vec{k},\ \vec{p},\ \vec{q}$ 
are constant $d$-dimensional vectors.
(If C=0, then this is simply a special case of the form discussed earlier.) 
Then the coefficient of $AB^2$ in (\ref{feq1}) is
%------------------------------------------------------------------------
\bea
&&\exp(p_s x^s)\left(p_\mu\frac{\pd}{\pd x^\nu}
-p_\nu\frac{\pd}{\pd x^\mu}\right)(p_\mu k_\nu-p_\nu 
k_\mu)\exp(p_rx^r+k_rx^r)\nonumber\\
&&= \exp(2p_rx^r+k_rx^r)\left(p^2k^2-(p\cdot k)^2\right)\,=\, 
0 \ . \label{sol1}
\eea
%------------------------------------------------------------------------
Similarly, the coefficient of  $AC^2$ in (\ref{feq1}) vanishes, 
provided
%------------------------------------------------------------------------
\be
q^2k^2-(q\cdot k)^2\,=\,0 \ . \label{sol3}
\ee
%------------------------------------------------------------------------
These conditions ensure that (\ref{feq2}) is also satisfied without 
any further condition. 
They have only nontrivial complex solutions in Euclidean space, 
but have real (or pure imaginary, as befits plane wave solutions) 
in Minkowski space. 
What is remarkable is that the cross term proportional to
$ABC$ in (\ref{feq1}) also vanishes if those two conditions 
(\ref{sol1}),(\ref{sol3}) are imposed.
The vanishing of this term is tantamount to the following 
condition;
%------------------------------------------------------------------------
\be
(p_\mu(q_\nu+k_\nu)-p_\nu(q_\mu+k_\mu))(q_\mu k_\nu-q_\nu 
k_\mu)+(q_\mu(p_\nu+k_\nu)-q_\nu(p_\mu+k_\mu))(p_\mu k_\nu
-p_\nu k_\mu)=0 \ .
\ee
%------------------------------------------------------------------------
Simplifying this equation by using the previous conditions, we have
%------------------------------------------------------------------------
\be
(p\cdot q)[(p+q+2k)\cdot k]-(p\cdot k)q^2 -(q\cdot k)p^2-2
(p\cdot k)(q\cdot k)\,=\,0 \ .
\label{mixedcon}
\ee
%------------------------------------------------------------------------
This is satisfied if the pair $(p,\, q)$ obeys a similar restriction to 
$(p,\, k)$ and $(q,\, k)$.
These restrictions enable (\ref{mixedcon}) to be expressed in the 
form
%------------------------------------------------------------------------
\be
(p\cdot q- |p||q|)(2|k|+|p|+|q|)\,=\,0 \ .  
\ee
%------------------------------------------------------------------------
which is clearly satisfied if $p\cdot q$ is replaced by $|p||q|$.
By a linear transformation, $k,\ p,\ q$ may be represented by 
three-dimensional vectors. 
$\vec k$ is then expressed as a linear combination,
%------------------------------------------------------------------------
\be
\vec{k}\,=\, 
\alpha\vec{p}+\beta\vec{q}+\gamma\vec{p}\times\vec{q} \ .
\ee
%------------------------------------------------------------------------
Taking scalar products of both sides with respect to $\vec{p}$ 
and $\vec q$ yields the required condition;
%------------------------------------------------------------------------
\be
\vec{p}\cdot\vec{q} =|p||q| \ .
\ee
%------------------------------------------------------------------------
With the metric $(+,\ -,\ -)$, the vectors may be parametrised by 
%------------------------------------------------------------------------
\bea 
\vec{k}&=&(\sinh{\theta},\ \cosh{\theta},\ 0) \ , \nonumber \\
\vec{p}&=&(a_1,\ a_2,\ a_1 \cosh{\theta}-a_2\sinh{\theta}) \ , \\ 
\vec{q}&=&(b_1,\ b_2,\ b_1 \cosh{\theta}-b_2\sinh{\theta}) \ . 
\nonumber
\eea
%------------------------------------------------------------------------
These vectors all satisfy the condition that the cross product of any 
pair are lightlike automatically. 
Further terms may also be added to both $u$ and  $v$, provided 
they fall within this parametrisation; all pairs have lightlike cross 
products, so the modes which fall within this class satisfy a 
superposition principle.
It is expected that the analysis can be extended at the cost of more 
technical complexity, to the case of three or more fields. 
The surprising result is that although the equations are non-linear, 
nevertheless a large class of plane wave solutions obey a 
superposition principle. 
Note further that the analysis extends to the situation where both 
fields are linear combinations of plane wave solutions; 
Suppose
%------------------------------------------------------------------------
\be
u= \sum_{j} A_j\exp(p_j\cdot x) \ , \qquad 
v= \sum_{j} B_j\exp(q_j\cdot x) \ .
\ee
%------------------------------------------------------------------------
Then apart from the trivial case where $u$ and $v$ are functionally 
related, the conditions for a solution of this type are simply
%------------------------------------------------------------------------
\be
(p_i\cdot p_j)^2-p_i^2p_j^2\,=\,(k_i\cdot 
k_j)^2-k_i^2k_j^2\,=\,(p_i\cdot k_j)^2-p_i^2k_j^2\,=\,0 \ , \ \ 
{\rm for} \ \forall\ (i,j) \ .
\ee
%------------------------------------------------------------------------
An important feature of a solution of this type is that the 
coefficients $A_j,\ B_j$ are all independent, and therefore may be 
treated as creation or annihilation operators in quantising the 
theory.
\par

%%%%%%%%%%%%%%%%%%%%%%%%%%%%%%%%
\section{Incorporation of mass}%
%%%%%%%%%%%%%%%%%%%%%%%%%%%%%%%%
A natural way to introduce a mass will be through the introduction 
of a term $M^{2n}\prod_1^n(\phi^j)^2$ into the Lagrangian. 
Then in the case of two fields in general dimension discussed in section
\ref{sqjl}, the conditions for a solution of the form of (\ref{sol}) require 
that $f,\ g$ are pure exponentials and  (\ref{cond}) is modified to 
%------------------------------------------------------------------------
\be 
(a_\rho a_\rho)(b_\sigma b_\sigma)\,-\,(a_\rho b_\rho)^2 \,=\, 
M^4 \ .
\ee
%------------------------------------------------------------------------
It seems that the incorporation of mass drastically reduces the 
variety of tractable solutions.
\par 

%%%%%%%%%%%%%%%%%%%%%
\section{Discussion}%
%%%%%%%%%%%%%%%%%%%%%
In the case of two fields, in section \ref{sqjl}, we have exhibited an 
arbitrary family of solutions to the second order equations of motion.
These solutions are complex in Euclidean space, but real in Minkowski 
space. 
On the other hand, in section \ref{afce}, we have demonstrated real 
solutions of the first order ansatz equation in Euclidean space.
This situation is reminiscent of the instanton construction,
as the self-dual ansatz for Yang-Mills also yields solutions 
which are real in Euclidean space. 
The solutions we have found are associated with spherical harmonics, 
labelled by two integers, which also brings to mind the fact that 
instantons also have a label, the instanton number.
\par

\newpage
%%%%%%%%%%%%%%%%%%%%%%%%%%%%
\section*{Acknowledgements}%
%%%%%%%%%%%%%%%%%%%%%%%%%%%%
D.B.F. would like to thank C.K. Zachos and R. Sasaki for useful 
discussions.\\
The work of T.U. is partially supported by the Daiwa Anglo-Japanese 
Foundation and the British Council. 
\par
\vspace{1.0cm}
%%%%%%%%%%%%%%%%%%%%%%%%%%%%%%%%%%%%%%%%%%%%%%%%%%%%%%%%%%%%%%%%%%

%%%%%%%%%%%%%%%%%%%%%%%%%%%%%%%%%%%%%%%%%%%%%%%%%%%%%%%%%%%%

\begin{thebibliography}{4}
\bibitem{bf1} L.M. Baker and D.B. Fairlie, `Companion Equations 
for Branes', \\ \ {\it J.~of~Math.~Phys.} {\bf 41} (2000) 4284-4292.\\
L.M. Baker and D.B. Fairlie, `Hamilton-Jacobi equations and Brane
associated Lagrangians',  {\it Nucl.~Phys.} {\bf B596}
(2001) 348-364.
\bibitem{dbf1} D.B. Fairlie, `Lagrange Brackets and $U(1)$ Fields',  
{\it Phys.~Lett.} {\bf B484} (2000) 333-336.
\bibitem{tu}D.B. Fairlie and T. Ueno, `Covariant formulation of 
Field Theories associated with $p$-Branes', DTP/00/93, 
{\bf hep-th/0011076} (2000), to appear in {\it J.~Phys.} {\bf A}.
\bibitem{ft} D.Z. Freedman and P.K. Townsend, `Antisymmetric 
Tensor Gauge theories and Nonlinear Sigma Models', 
{\it Nucl.~Phys.} {\bf B177} (1981) 282.
\bibitem{cos} C.K. Zachos, private correspondence.\\
\ R. Jackiw, `Descendants of the Chiral Anomaly',
{\bf hep-th/0011275} (2000).\\
\ R. Jackiw, `Collaborating with David Gross; Descendants of the 
Chiral Anomaly', {\bf hep-th/013017} (2001).
\bibitem{cosmas}C. Zachos, D. Fairlie and T. Curtright, `Matrix 
Membranes and Integrability', {\bf hep-th/970942}, in 
{\it Supersymmetry and Integrable Systems} (Springer Lecture 
Notes in Physics, ed. by H. Aratyn, 1997).
\bibitem{ward} R. Ward, {\it Phy.~Lett.} {\bf B234} (1990) 81.
\end{thebibliography}
\end{document}